\documentclass[useAMS,usenatbib]{mn2e}
\usepackage{graphicx}
\usepackage{float}
\usepackage[authoryear]{natbib}
\usepackage{color}
\title[Observations and 3D HD models of PNe]{Observations and
3D Hydrodynamical models of planetary nebulae with Wolf Rayet type central stars \thanks{Based upon observations carried out at the Observatorio Astron\'omico Nacional on the Sierra San Pedro M\'artir (OAN-SPM), Baja California, M\'exico.}}
\author[Rechy-Garc\'ia, J. S., et al]{Rechy-Garc\'ia, J. S. $^{1}$\thanks{E-mail: jrechy@astro.unam.mx}, Vel\'azquez, P. F. $^{2}$, Pe\~na. M.$^{1}$, Raga, A. C.$^{2}$\\
\\
$^{1}$Instituto de Astronom\'ia, Universidad Nacional Aut\'onoma de M\'exico, Apdo. Postal 70264, 04510, Ciudad de M\'exico, M\'exico. \\
$^{2}$Instituto de Ciencias Nucleares,  Universidad Nacional Aut\'onoma de M\'exico, Apdo. Postal 70543, 04510, Ciudad de M\'exico, M\'exico. }

\begin{document}

\date{}

\pagerange{\pageref{firstpage}--\pageref{lastpage}} \pubyear{2002}

\maketitle

\label{firstpage}

\begin{abstract}
We present high-resolution, long-slit spectroscopic observations of two
planetary nebulae with [WC] central stars located near the galactic
bulge, M\,1-32 and M\,3-15. The observations were obtained with the
2.1-m telescope at the Observatorio Astron\'omico Nacional, San Pedro
M\'artir. M\,1-32 shows wide wings on the base of its emission lines
and M\,3-15 has two very faint high-velocity knots.  In order to model
both PNe we built a three-dimensional model consisting of a jet
interacting with an equatorially concentrated slow wind, emulating the
presence of a dense torus, using the Yguaz\'u hydrodynamical code.
From our hydrodynamical models, we obtained position-velocity
(PV) diagrams in the [N\,{\sc ii}]$\lambda$6583 line for comparison with the
observations. We find that the spectral characteristics of
M\,1-32 and M\,3-15 can be explained with the same physical model -a
jet moving inside an AGB wind- using different parameters (physical
conditions and position angles of the jet). In agreement with our model and observations, 
these objects contain a dense torus seeing pole-on and a bipolar jet escaping  thorough the poles. Then we propose to classify this kind of objects as spectroscopic bipolar nebulae, although
they have been classified morphologically as
 compact, round, or elliptical nebulae or with "close collimated lobes".

\end{abstract}


\section{Introduction}

Planetary nebulae (PNe) represent a highly evolved stage of
low-intermediate mass stars (1-8 M$_\odot$). Several thousand  objects are known
in the Milky Way and chemical abundances have been determined in a
large number of them which has led to classify them as Type I (N and
He-rich young objects), Type II (objects in the galactic disk with
peculiar velocities lower than 60 km s$^{-1}$), Type III (objects with
peculiar velocities larger than 60 km s$^{-1}$), and Type IV
\citep[old objects belonging to the  extreme Population II, see][]{Peimbert1990}.
Thus, PNe of different types belong to different populations in the galaxy.

In addition, PNe are classified by their morphology: they can be
round, compact, elliptical, bipolar, or point-symmetric nebulae
\citep[see for example,][]{Balick1987,Manchado1996}. More
  recently, \citet{SahaiMorris2011} proposed a more detailed morphological
  classification for young PNe, based on deep, high-resolution imaging from HST.
  We will refer to this classification later on. However, in many cases 
  it is difficult to decide which is the real morphology
of a PN, because we need to consider different aspects. For example
the projection effects of a three dimensional object on the plane of
the sky, the exposure time dependence on the object morphology, or
observations in lines of different ions give different morphologies
due to the stratification in the nebula. Despite these complications,
the study of morphology plays an important role for understanding the
ejection processes, the presence of magnetic fields and even the
presence of a binary central star.

Approximately 80\% of PNe are asymmetric \citep[]{Parker2006, Douchin2013},
showing ellipsoidal, bipolar or multipolar morphologies, displaying
tori, jets, knots, etc. How can the formation of such components be
explained? \citet{Pascoli1992} proposed that the magnetic field
can produce a bipolar morphology in
PNe. Also \citet{GarciaSegura1999} suggested that
asymmetries can be due to magnetic fields and stellar
rotation.
\citet{SokerHarpaz1992} had already shown that even if
the magnetic field works for shaping asymmetric PNe, a binary
companion is necessary.
The notion that magnetic fields alone are insufficient for shaping 
bipolar morphologies has been subsequently reinforced 
by \citet{Soker2006}, \citet{Nordhaus2007}, and \citet{GarciaSegura2014}. 

The conception of a binary system as the shaping mechanism 
of bipolar and multipolar PNe has several decades 
\citep[]{Bond1978, Livio1979, SokerLivio1994, DeMarco2009}.
\citet{Bond2000} showed that there is a considerable circumstantial evidence about the role played by close-binary systems (or post-common envelope systems, post CE ) in the formation and shaping elliptical or bipolar planetary nebulae. \citet{Tocknell2014} found that  post CE central stars often have jets.
 \citet{DeMarco2008} showed that at least 10\%--15\% of PNe have binary systems with
short orbital periods ($<$3 days), giving strong support to the binary hypothesis as shaping mechanism. Several more binary systems in PNe have been reported by \citet{Miszalski2009}. 

Bipolar PNe
can be formed by two jets which are
ejected by one of the components of a binary system \citep{Morris1987, SokerRappaport2000}.
\citet{SahaiTrauger1998}  \citep[see also][]{BalickFrank2002} showed that
these jets play a primary role in forming bipolar proto PNe (PPNe).  \citet{GarciaArredondo2004} carried out a 3D hydrodynamical simulation of the interaction between an AGB wind and a jet, showing that bipolar morphologies are obtained for the strong jet case (when the jet momentum is larger that the AGB wind momentum). \citet{Dennis2008} showed that ``jetlike" outflows in PPN can be formed by the launching of clumps or bullets from one of the components of the binary system.
Employing the
hypothesis of a binary system in the centre of the nebula,  \citet{Riera2014} and \citet{Velazquez2014} modelled the morphology and proper motions of the PPNe CRL 618 by means of a precessing jet with a time-dependent ejection velocity, 
obtaining a good agreement with observations. We will use these ideas in this work.

\smallskip

A special group of PNe (around 10\%) has a central star which displays
strong and very
wide emission lines, evidence that they are suffering atmospheric
instabilities with large mass loss. These stars show spectra similar
to the massive Wolf-Rayet stars of the C sequence and they have been
classified as [WC].
Their atmospheres are deficient in H, showing mainly He, C and
O \citep{Koesterke2001}.  PNe with central stars of the  [WC] type 
(hereafter [WC]PN) have  larger
expansion velocites and present more turbulence than PNe with normal central
stars \citep{Medina2006}. \citep{Pena2013} found that the [WC]PNe are
more concentrated towards the galactic thin disk (height smaller than
about 400 pc from the disk), compared to PNe with normal central
stars, most of which are distributed up to 800 pc from the disk. This
has been interpreted as evidence that [WC]PNe belong to a younger population than
PNe with normal stars.

In this work  we aim to analyze the kinematical behavior of two [WC]PNe, named 
M\,1-32 and M\,3-15, which show evidence of having jets at high velocities and 
to relate their kinematics with other nebular characteristics. 

\begin{table*}
\centering
 \begin{minipage}{130mm}
  \caption{Main characteristics of M\,1-32 and M\,3-15}
\begin{tabular}{llllllll}
\hline \hline 
PN G & Name  & Spectral & $V_{hel}$&$D$ &C/H$^a$&N/H$^a$ & O/H$^a$\\
& &Type& km s$^{-1}$&kpc&&\\
\hline
006.8$+$04.1&M\,3-15&[WC\,4]&  96.9$\pm$0.8&6.82$\pm$1.365& 8.85$\pm$0.13 & 8.33$\pm$0.10&9.18$\pm$0.08\\ 
011.9$+$04.2&M\,1-32&[WO\,4]\,pec&-86.4$\pm$0.8&4.79$\pm$0.959&9.75$\pm$0.09&8.44$\pm$0.06&9.11$\pm$0.08\\
\hline
\multicolumn{4}{l}{$^a$ In 12 + log X/H}  (see \citet{GarciaRojas2013})
\end{tabular}
\label{tab:charac}
\end{minipage}
\end{table*}

 M\,1-32 (PN G 011.9+04.2) has a [WO\,4]\,pec central star
 \citep[]{AckerNeiner2003, WeidmannGamen2011}
 and presents some very interesting
 characteristics, some of which are listed in Table \ref{tab:charac}.  
 It is located towards the galactic bulge, showing a
 high heliocentric radial velocity of $-$86.4 km s$^{-1}$
 \citep{Pena2013}.  Its heliocentric distance is 4.79
 kpc \citep{Stanghellini2010}. It is a moderate Peimbert Type I PN (with
a N/O abundance ratio of 0.50
 and a He/H ratio of 0.126) and shows a large C-enrichment,
 with a C/O abundance ratio of 4.3 \citep{ GarciaRojas2013}. Evolution
 models for single stars by \citet{Karakas2010} indicate that such a
 nebular C-enrichment should have been produced by a central star with
 an initial mass larger than 3 M$_{\odot}$. This high initial stellar
 mass can also explain the high N/O abundance in the nebula.
 Morphologically it has been classified as an elliptical nebula \citep{Stanghellini2010} but, from
 high-resolution spectra \citet{Medina2006} found that the nebular
 lines present intense narrow profiles with faint wide wings. In
 addition¼, from spectroscopic analysis it is found that the nebular
 morphology of M\,1-32 is highly asymmetric, showing a dense,
almost face-on torus
with low expansion velocity, and also knots and
 jets at high velocities of $\pm$ 200 km s$^{-1}$. There is also
 evidence of a possible external envelope \citep{ AkrasLopez2012}. \\

 PN G 006.8+04.1, also known as M\,3-15, has a [WC\,4] central star
 \citep[]{AckerNeiner2003, WeidmannGamen2011} and 
 it has been classified as compact  \citep{Lopez2012}.  It is
 located towards the galactic bulge, showing a high heliocentric
 radial velocity of 96.9 km s$^{-1}$ \citep{Pena2013}. Its
 heliocentric distance is 6.82 kpc \citep{Stanghellini2010}.  In Table
 \ref{tab:charac} we list some of its characteristics.
 \citet{Medina2006} found that the line profiles show a narrow
 component, with an expansion velocity of 20 km s$^{-1}$ and two faint
 knots at velocities of $\pm$ 90 km s$^{-1}$ \citep[see
 also][]{AkrasLopez2012}. The total nebular abundance ratios in this
 object are C/O = 0.467, N/O = 0.14 y He/H = 0.107
 \citep{GarciaRojas2013}; therefore, differently from M\,1-32, this
 object is not a Type I PN and it does not show C-enrichment.

\begin{figure}
\begin{center}
\includegraphics[width=0.85\linewidth]{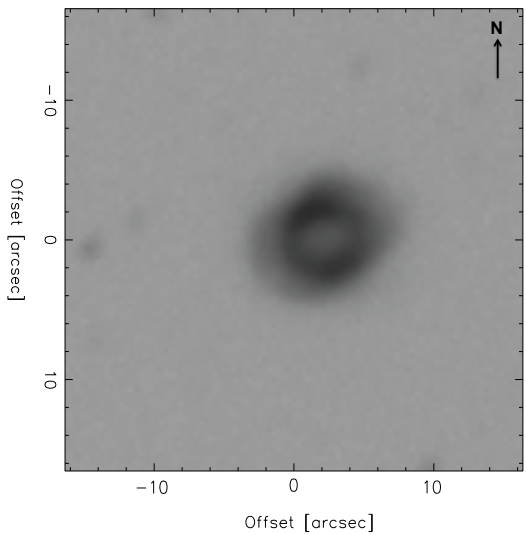}
  \caption{ H$\alpha$ image of PN M\,1-32, obtained that OAN-SPM using MES spectrograph. North
  is up and East to the left. The exposure time is 100 s. We observe a face-on torus and a possible external envelope.}
  \label{fig:M132image}
 \end{center}
\end{figure}

 In the recent morphological classification scheme proposed by \citet{SahaiMorris2011},
   based on HST imaging, they classify M3-15 as a ``L,c bcr(c) ps(s)''
   nebula, meaning that it has close collimated lobes, with a close
   ``barrel-shaped" central region, and the overall geometric shape of
   the lobes is point-symmetric \citep[see Figure 6 of][]{SahaiMorris2011}.
   There are no high-resolution images
   available for M1-32, so that \citet{SahaiMorris2011} classification
   cannot be directly applied, but in Figure \ref{fig:M132image} we present an image in H$\alpha$ of M1-32 obtained by us
   which clearly shows that the shape for this PN is a ``face-on torus''.

In this work, we use high resolution spectra to analyze the kinematical
behavior of the gas in these two [WC]PNe, and by using a
hydrodynamical model we investigate if the morphological structure of
M\,1-32 and M\,3-15 can be produced by the interaction between a jet
and a dense torus. We compare the
position-velocity diagrams obtained from the observations with those
predicted by the hydrodynamical models.

The paper is organized as follows. In \S 2, the spectral observations
are presented. In \S 3, we describe the hydrodynamical model which has
been calculated with the code Yguaz\'u \citep{Raga2000}. In \S 4 we
discuss the results for each object and in \S 5 we summarize
the main conclusions.

\section{Observations }

 High-resolution spectra for both planetary nebulae
  were obtained with the Manchester Echelle
 Spectrometer \citep[MES,][]{Meaburn2003} attached to the 2.1-m telescope of the Observatorio
 Astron\'omico Nacional at San Pedro M\'artir, B.C, Mexico(OAN-SPM). The
MES is a long-slit echelle spectrometer, which 
uses narrow-band filters to isolate the orders containing the emission lines
of interest. In our case, the filters used   isolated the order corresponding to H$\alpha$ 
(this order includes the H$\alpha$, [N\,{\sc ii}]$\lambda$6548 and [N\,{\sc ii}]$\lambda$6583 emission  lines) and the order isolating the [O\,{\sc iii}]$\lambda$5007 
emission line.  In this paper, we only show the behavior of the
[N\,{\sc ii}]$\lambda$6548 line because the other 
 lines show the same structure. 

The MES  was equipped with a Marconi 2 CCD detector, which has a pixel size of 13.5 $\mu$m. A binning of 2$\times$2 was applied and, using the secondary mirror f/7.5 which provides a plate scale of 13.2 $''$/mm, the plate  scale on the  detector was 0.356 $''$/pixel. The  slit width 
used was 150 $\micron$ (equivalent to $2''$ on the sky) and it was oriented north-south. The  spectral resolution obtained is of 11 km s$^{-1}$. 

Observations were obtained for three different regions of the object (see
 Figs. \ref{fig:M1-32} and \ref{fig:M3-15}). Immediately after every science observation, 
exposures of a Th-Ar lamp were acquired for wavelength calibration.  The internal precision 
of the  lamp calibrations
 is better than 1.0 km s$^{-1}$. Data reduction  was performed using the standard
  processes for the MES, with IRAF\footnote{IRAF is distributed by the National Optical Astronomy Observatories, which is operated by the Association of Universities for Research in Astronomy, Inc., under contract to the National Science Foundation.} reduction packages. In Table~\ref{tab:log} we present the log of the observations.
 
\begin{table}
\centering
 \begin{minipage}{140mm}
   \caption{Log of observations}
\begin{tabular}{llllc}
\hline \hline
PN G & Name  & Date& slit pos., PA & Exp.\\
 & & & & time (s)\\
\hline
006.8+04.1&M\,3-15&05/08/2011&East, 0$^{\circ}$&   600\\ 
&&&Centre, 0$^{\circ}$&1200\\
&&&West, 0$^{\circ}$&1200\\
011.9+04.2&M\,1-32&04/08/2011&East, 0$^{\circ}$&1200\\
&&&Centre, 0$^{\circ}$&1200\\
&&&West, 0$^{\circ}$&  600\\
\hline
\end{tabular}
\label{tab:log}
\end{minipage}
\end{table}

\subsection{PV diagrams}
\label{sec:space}
 
\begin{figure}
\includegraphics[width=1.1\linewidth]{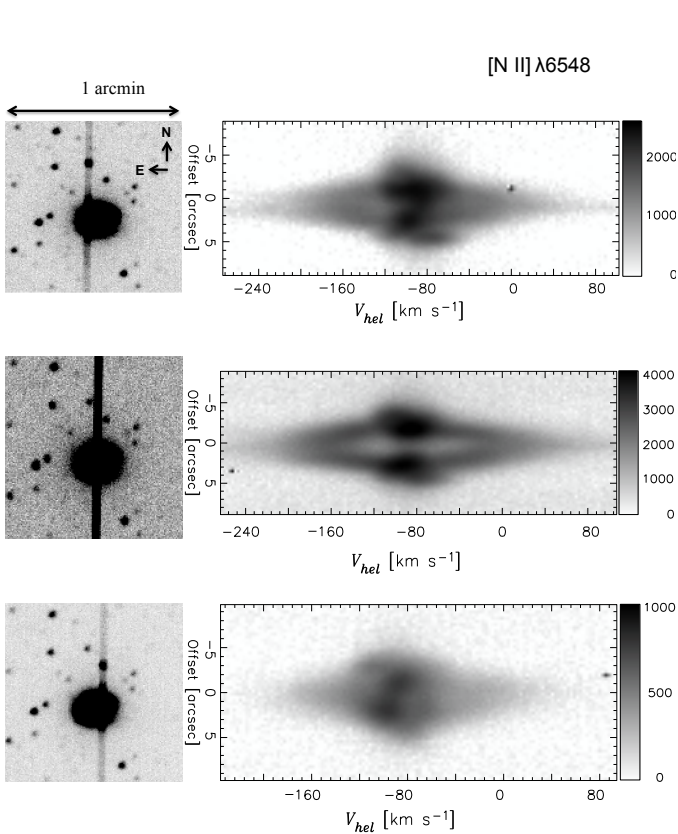}
  \caption{ Position-velocity diagrams of M\,1-32 in the [N\,{\sc
        ii}]$\lambda$6548 line,  for three position of the
    slit (shown in the left frames). The Eastern
    and Western slits are located at about 2.8 arcsec  and 2.5 arcsec
    (respectively)  from
    the centre of the object. The size of the images
     is 1.0 $\times$ 1.0 arcmin. In the diagrams to the right, the intense knots correspond to the
    toroidal component. The wide faint components come from a bipolar
    jet. The grey scale to the right is the intensity in counts. The ratio between the brightest and faintest regions is of the order of 30.}
  \label{fig:M1-32}
\end{figure}

PV diagrams for each PN were obtained with the WIP software
\citep{Morgan1995}.  In Fig. \ref{fig:M1-32} the PV diagrams for
M\,1-32 are shown for the slit located in three different positions
(East, Centre, and West, as it is shown in the left side of this figure). 
In the horizontal axis the heliocentric
radial velocity is represented and the vertical axis shows the spatial
direction.  The emission line shown in these spectra is [N\,{\sc
  ii}]$\lambda$6548. In the central position diagram two bright knots (at a
heliocentric systemic velocity of $-$86.4 km s$^{-1}$) and wings at
high velocities are visible. These wings have velocities of $\pm$180 km s$^{-1}$, relative to the systemic velocity. 

\begin{figure}
 \includegraphics[width=1.1\linewidth]{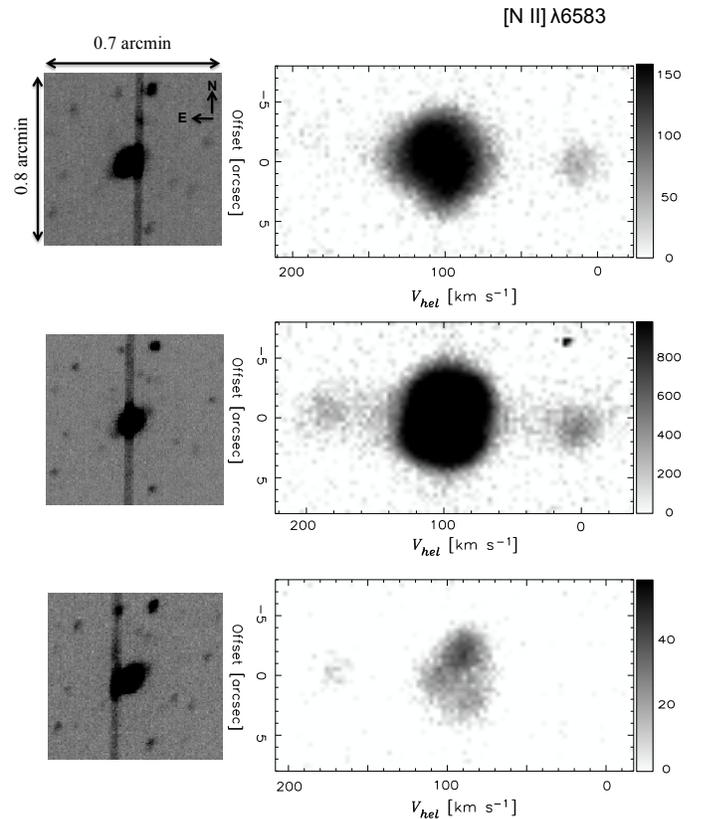}
 \caption{Position-velocity diagrams of M\,3-15 in the [N\,{\sc
       ii}]$\lambda$6583 line, for three slit positions (shown in
   an image in the left frames). the size of the images
     is 0.8 $\times$ 0.7 arcmin. The Eastern
    and Western slits are located at 2.0 arcsec from
    the centre of the nebula. The intense central emission corresponds to a toroidal
   component which is more compact than in the case of M\,1-32. The
   wide faint components correspond to a bipolar jet. The grey scale to the right is the intensity in counts. The ratio between the brightest and faintest regions is of the order of 30.}
\label{fig:M3-15}
\end{figure}

Fig. \ref{fig:M3-15} shows the PV diagrams for M\,3-15, obtained for
three slit positions. 
As for the case of M\,1-32, the emission line shown in these spectra is [N\,{\sc ii}]$\lambda$6548. In the central slit
 position a bright condensation (at a systemic velocity of 96 km s$^{-1}$) and jets,
ending in knots at $\pm$90 km s$^{-1}$ (relative to the systemic velocity) are seen.
In the East and West slit positions we only observe one side of the
jet and one knot.

The observations are not calibrated in flux, but we have added a scale in counts.
 In the observational PV diagrams, the ratio between the brightest and faintest 
 regions is of the order of 30.

\section{Numerical simulations}
\subsection{The hydrodynamical model and the Yguaz\'u code}

We have calculated hydrodynamical models using the code Yguaz\'u. This is a
3D adaptive code which solves the gasdynamic equations using the  `flux vector splitting" algorithm of \citet{VanLeer1982}. 
Together with these
equations, it also integrates a system of equations for atomic/ionic
species, which are used for obtaining the
cooling function \citep{Raga2000}. For this simulation, we have used
the total abundances of C, N, O and S (relative to hydrogen), obtained
by \citet{GarciaRojas2013}.  

From the observations it is evident that both objects have a dense
torus and that the high velocity wings flow through the poles of the torus. 
Therefore, in order to calculate a hydrodynamical model for these objects, we have considered a circumstellar medium
which was previously swept up by a dense low-velocity wind of an AGB star. Afterwards, a second
outflow is launched by the central star. Different types of second outflow were modeled: an isotropic wind,
an anisotropic wind (with anisotropical distributions in density or velocity), a cylindrical jet and a conical jet.

We finally found that the  model that best describes the PV diagrams observed for
 M\,1-32 and M\,3-15 nebula consists of two components: 
a dense torus and a bipolar system of conical jets.
However, this model does not reproduce the
bright regions at radial velocities close to zero, which are observed in the PV diagrams. In order
to reconcile observations with models, it was necessary to include the
effect of a central star of the nebula, which would ionise the
surrounding medium. Due to the fact that the Yguaz\'u hydrodynamical code does not solve the radiative
transfer equation, we have included the photoionisation effects of
a central source by keeping  H, He, C, O, and N at least singly ionised.
 
\bigskip

\subsection{Initial setup}
Following \citet{Mellema1995}, at the beginning of our simulations, we have imposed on all
of the computational domain an AGB wind which has a density
distribution given by

\begin{equation}
\rho (r, \theta) = \rho_{0} g (\theta) (r_{0}/r)^{2},
\label{rhor}
\end{equation}

\noindent where $\rho_0 = \dot M / (4 \pi r_{0}^2 v) $ is the initial density of the AGB slow wind, with mass loss rate $\dot M$ 
and  a terminal velocity $v$ of 20 km s$^{-1}$. The angular dependence of the wind is given by the
parametrized function
\begin{equation}
  g(\theta) = 1 - A \left[\frac{1-\exp(-2 B \cos^{2} \theta)}{1-\exp(-2 B)}\right],
  \label{g}
\end{equation}
where $A$ determines the ratio between the density at the equator
and the pole; and $B$ determines the way the density varies from
the equator to the pole, $r$ is the distance from the central star and
$\theta$ is the polar angle (0$^{\circ}$ at the pole and 90$^{\circ}$
at the equator).
In the equation \ref{rhor}, $r_0$ is a reference radius which is equivalent to
5 pixels of the grid at maximum resolution. The values of $r_0$ employed in our simulations are listed
in table \ref{tab:hydro-model}.

\begin{figure}
\centering
   \includegraphics[width=1.0\linewidth]{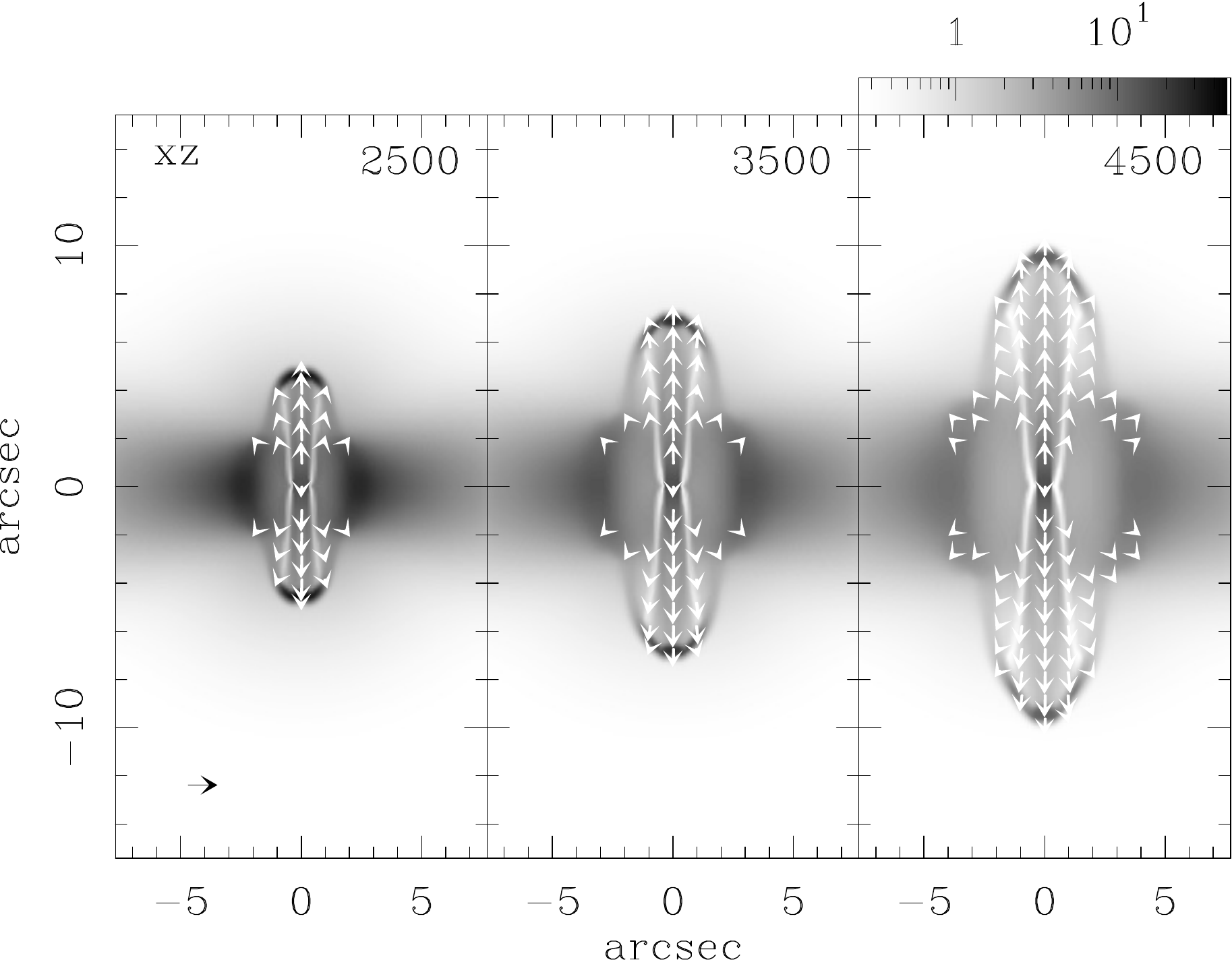}
 \includegraphics[width=1.0\linewidth]{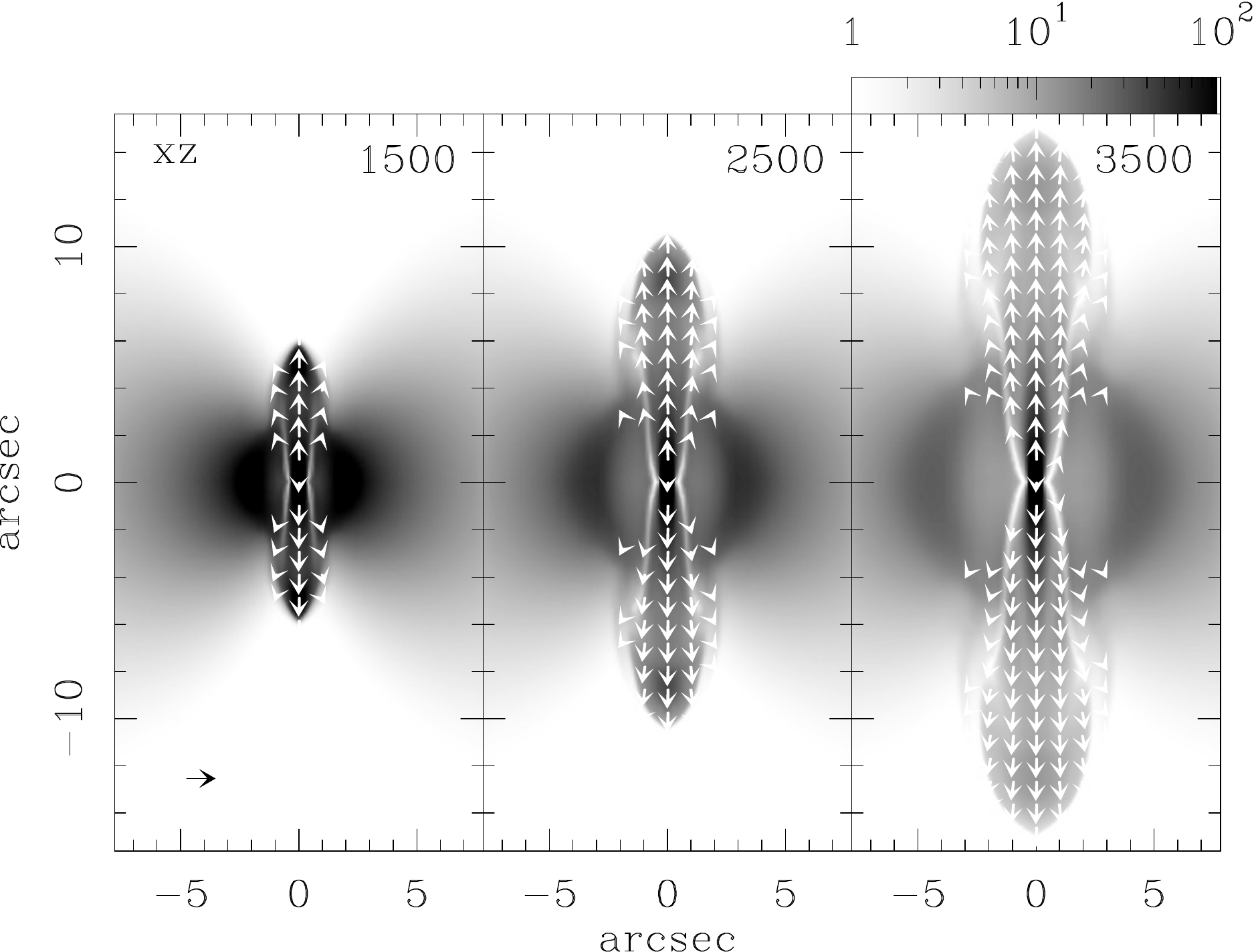}
  \caption{Top: Electron density distribution map ($xz-$projection),
    corresponding to the model for M\,1-32 at different evolutionary times in yr. 
    Bottom: the same map for M\,3-15. The grey scale (given by the bars on the
    top of the right hand side frames) is given in
    units of particles $\mathrm{cm}^{-3}$. Both axes are given in units of
    arcseconds, considering a distance of 4.8 kpc and 6.8 kpc for M\,1-32 and M\,3-15, respectively.
    The arrows show the velocity field of the jet. The small arrows at the bottom of
    the left frames correspond to 200 km s$^{-1}$ .}
  \label{fig:model}
\end{figure}

For the case of PN M 1-32 we have
chosen $A=0.9$ and $B=5$, and for M 3-15 we have chosen $A=0.99$ and $B=0.1$.
For both models,  a temperature of 1000 K was adopted for
the AGB wind.  In Table \ref{tab:hydro-model} we list the values for $\dot M$,
the initial jet radius r$_{j}$ (which was set equal to $r_0$) , the jet velocity v$_{j}$, the semi-aperture of the conical jet $\alpha$, and the size of the
computational domain employed for our PN models.

\begin{table*}
\centering
 \begin{minipage}{120mm}
  \caption{Input for hydrodynamical simulations}
\begin{tabular}{lllccccc}
\hline \hline
Name & Computational domain&A&B& r$_{j} = r_{0}$ & $v_{j}$ &$\alpha$ &$\dot{M}$\\
& $x-$ , $y-$, and $z-$axis, (cm) &&& (cm) & (km s$^{-1}$)& (degrees)& ($\mathrm{M_{\sun}\ yr^{-1}}$) \\
\hline
M\,1-32 & (2.25, 2.25, 4.5)$\times$ 10$^{18}$  &0.9&5&4.4 $\times$ 10$^{16}$ & 150 & 5 & 5 $\times$ 10$^{-6}$\\
M\,3-15 & (1.57, 1.57, 3.15)$\times$ 10$^{18}$  &0.99&0.1&3.08 $\times$ 10$^{18}$ & 140 & 5 & 6 $\times$ 10$^{-6}$\\
\hline
\end{tabular}
\label{tab:hydro-model}
\end{minipage}
\end{table*}

\section{Results and discussion}

Fig. \ref{fig:model} shows the time evolution of the
  electron density distribution of the models for M\,1-32 and M\,3-15. Also, this figure shows
  the velocity field of the flow. A clear bipolar morphology results from the presence of the
  conical jets. This outflow expands into a latitude-dependent AGB
  wind, with high densities at the Equator.
In order to compare the observations (Figs. \ref{fig:M1-32} and
  \ref{fig:M3-15}) with the hydrodynamical model, we also
  obtained PV diagrams from the models.

\begin{figure*}
\centering
	
	\includegraphics[width=0.45\linewidth]{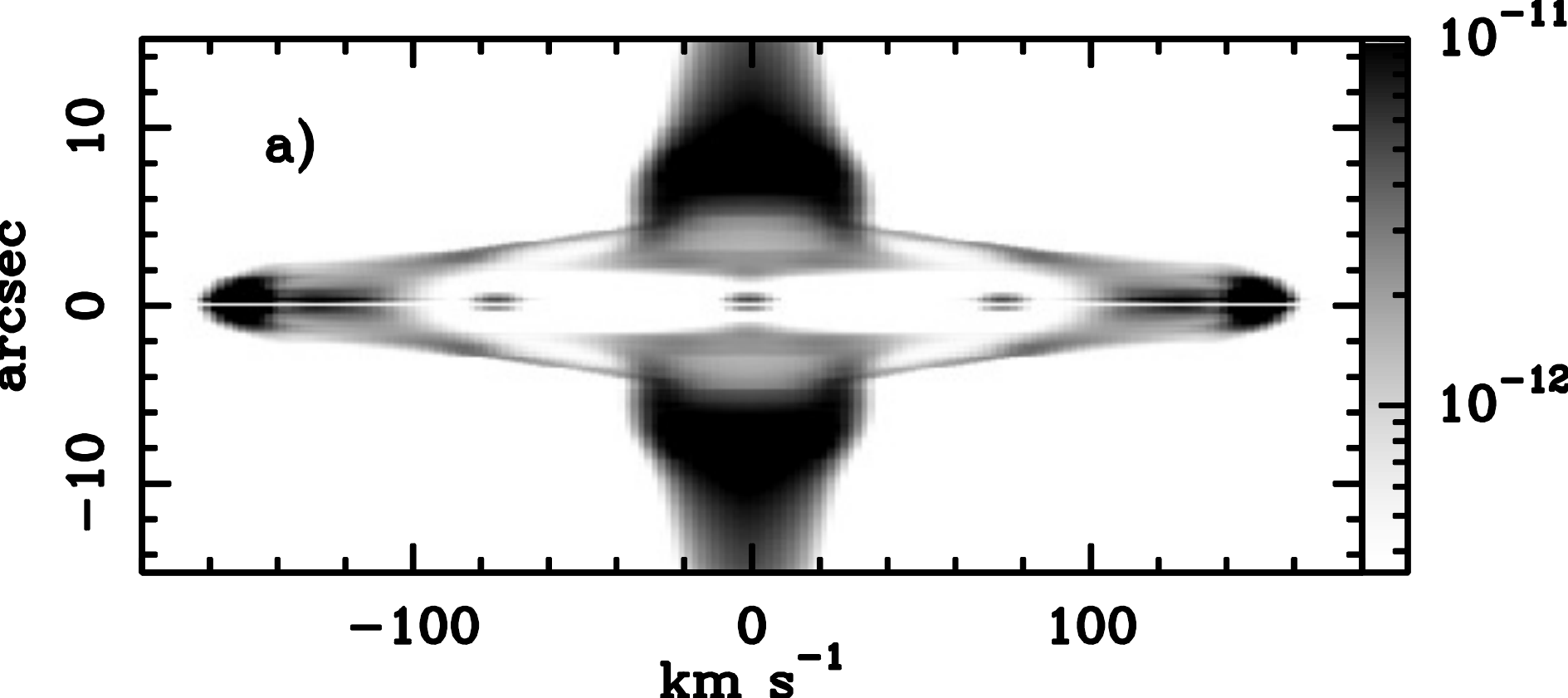}
	\includegraphics[width=0.45\linewidth]{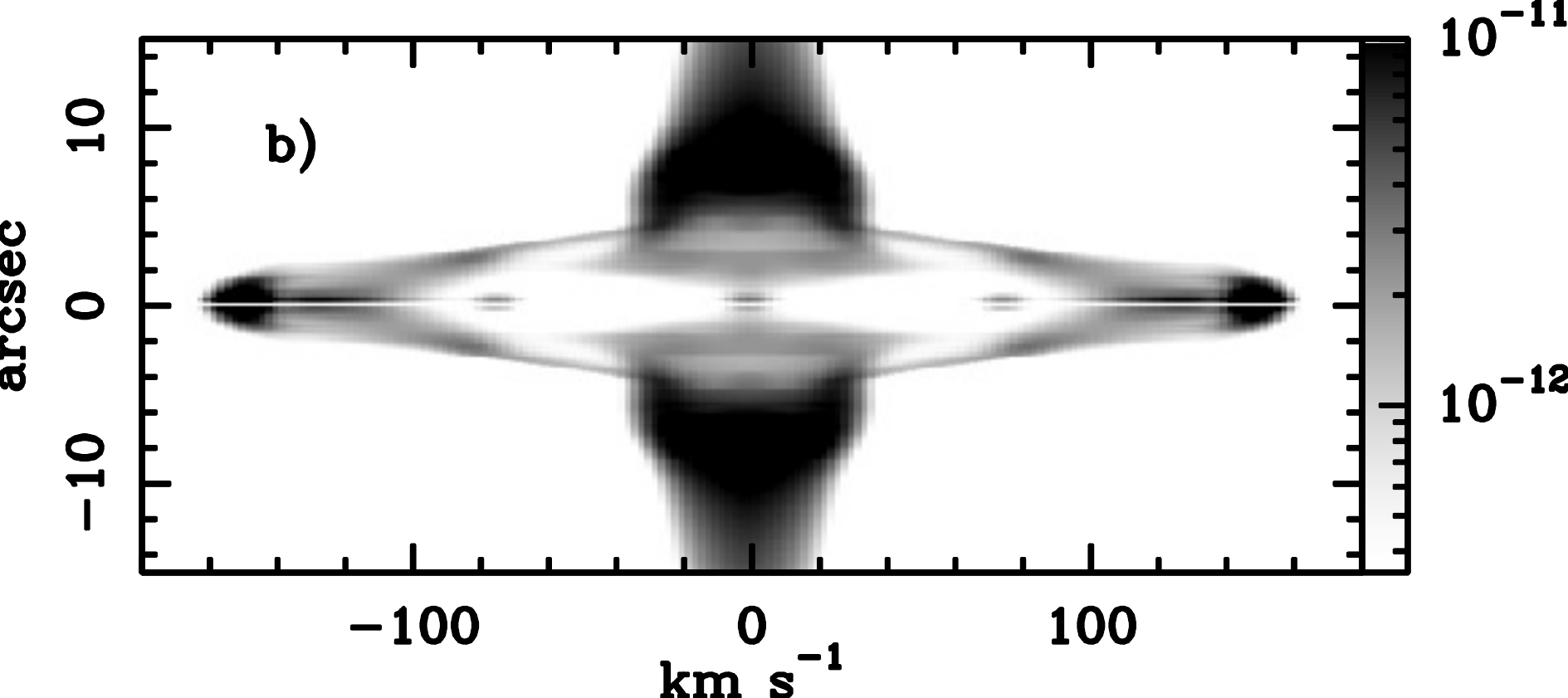}
	\includegraphics[width=0.45\linewidth]{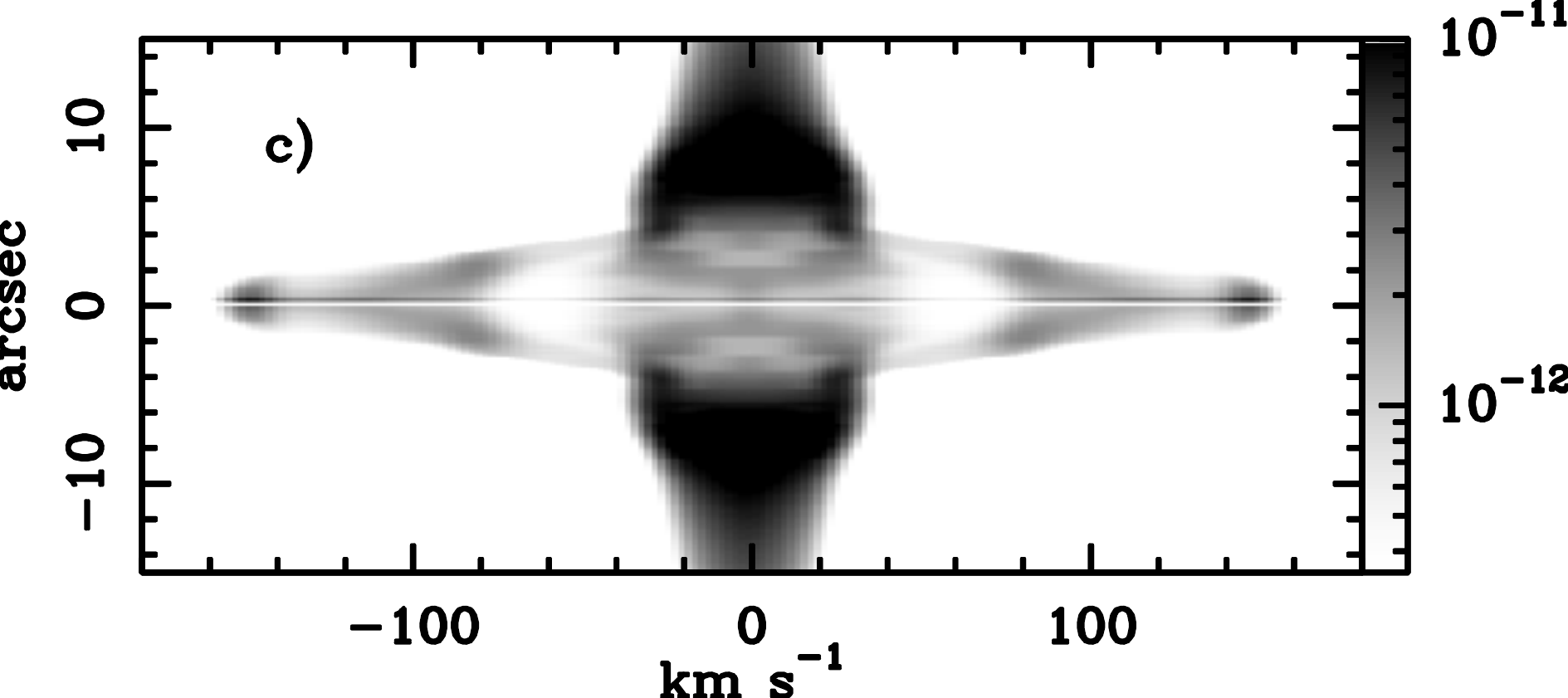}
	\includegraphics[width=0.45\linewidth]{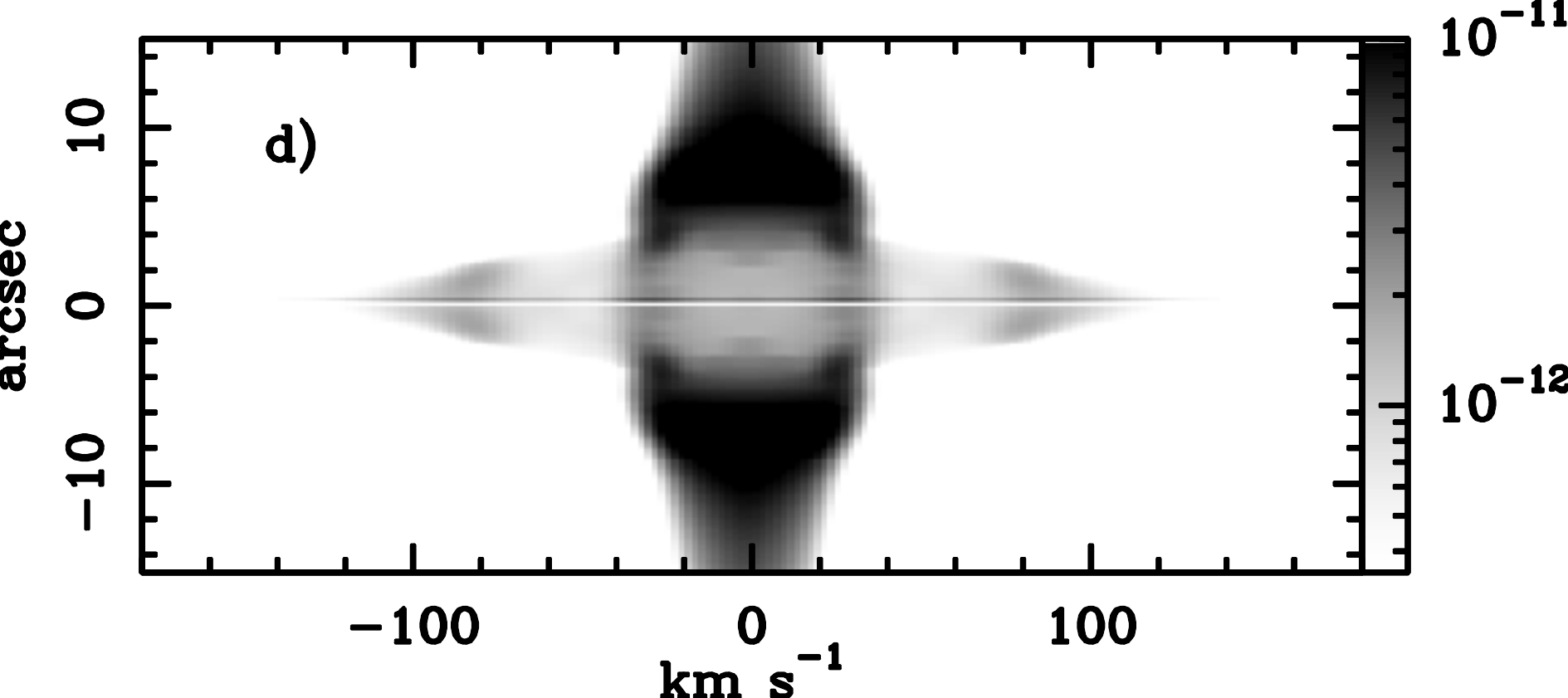}
	
	\caption{Synthetic position-velocity diagrams of the [N\,{\sc
        ii}]$\lambda$6583 line, obtained from the hydrodynamical
          model for M\,1-32. a) PV diagram for the slit located 
          on the axis of the model.  b) Slit at 1 arcsec from the
          axis. c) Slit located at 2 arcsec from axis. d) Slit at
          3 arcsec from the axis. The horizontal and vertical axes represent the velocity in
          km s$^{-1}$ and the spatial direction in arcsec,
          respectively. The logarithmic grey scale is given in units of
          erg cm$^{-3}$ sr$^{-1}$. }
  \label{fig:pv-M1-32}
\end{figure*}

  The simulation for M\,1-32 was carried out until an integration time
  of 4500 yr.  Synthetic PV diagrams (for the [N\,{\sc
    ii}]$\lambda$6583 emission line) were obtained considering several
  slits located at different distances from the symmetry axis of the
  object. Due to the fact that the observed PV diagrams of this PN are
  symmetrical on both sides of the axis, we only display the PV
  diagrams obtained for slits located to the right of the PN source
  (the same results are obtained if we move the slit to the left).
  These PV diagrams are shown in Fig. \ref{fig:pv-M1-32}.

The bright central knots come from the torus, which has a total diameter of
$20''$ and it is at a velocity of 
0 km s$^{-1}$, while the wings (associated with the jet) show
velocities ranging up to $\pm$150 km s$^{-1}$. In these synthetic
PV diagrams, for slits located farther away from the
centre, the inner zone of the torus is filling up, due to the fact that the slits approach the edge of the PN.
In addition, in PV diagram for Fig. \ref{fig:pv-M1-32} d) only part of the jet is seen, giving the appearance that the wings are moving at slow velocities. 

For M\,1-32, our synthetic PV diagrams (Fig. \ref{fig:pv-M1-32}) are
in good agreement with the observational ones (Fig. \ref{fig:M1-32}),
especially the synthetic PV diagrams labeled a), c) and d) corresponding to slits
located on the axis, at $2''$, and at $3''$ from the centre, respectively.  We
reproduce the central knots corresponding to the dense torus, and the
wide wings associated with the jet material.

In our model for
M\,1-32, the jet axis is nearly pole-on, i.e., at 0$^{\circ}$ with
respect to the line of sight. The synthetic PV diagrams show fast
bipolar outflows at velocities of 160 km s$^{-1}$, with shapes that are similar
to the observations (Fig. \ref{fig:M1-32}), and a dense torus
which expands at velocities of approximately 25 km 
s$^{-1}$ with respect to the centre of the nebula (also in qualitative agreement with the observations).

The fact that we assume that all of the material in the computational
domain is at least singly ionized results in strong emission from
the bowshocks at the leading edge of the conical jets. This emission
produces the bright knots at radial velocities of $\pm 160$~km~s$^{-1}$
in the PV diagrams corresponding to slit position 
a) and b) (Fig. \ref{fig:pv-M1-32}).

\begin{figure}
\centering

\includegraphics[width=1.0\linewidth]{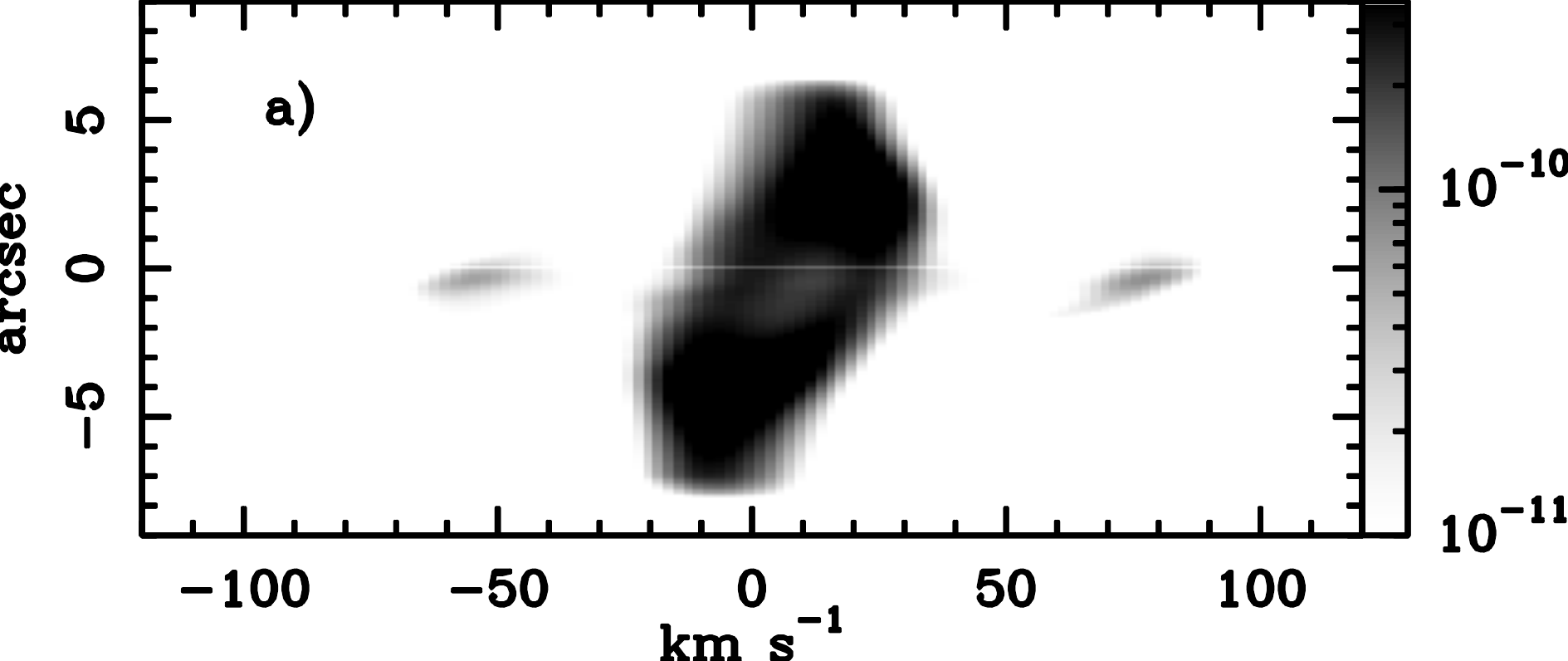}
\includegraphics[width=1.0\linewidth]{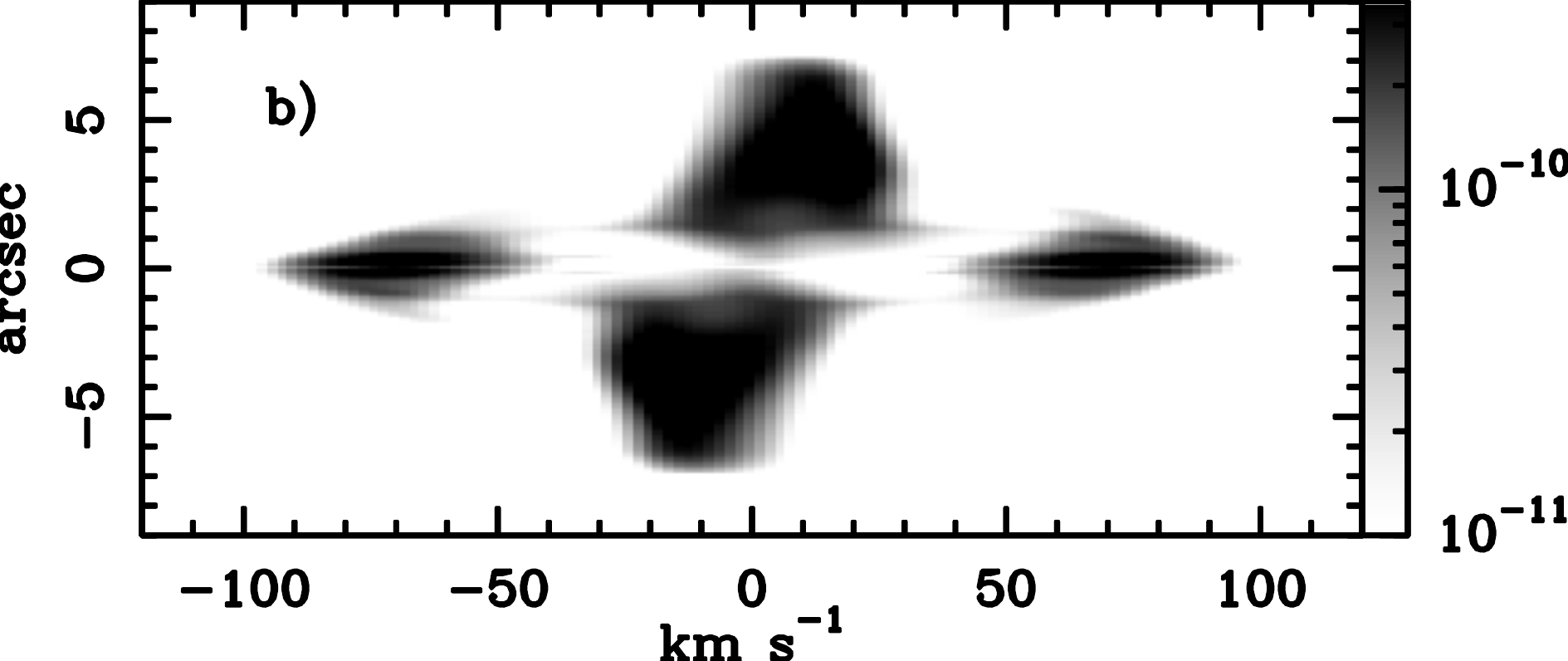}
\includegraphics[width=1.0\linewidth]{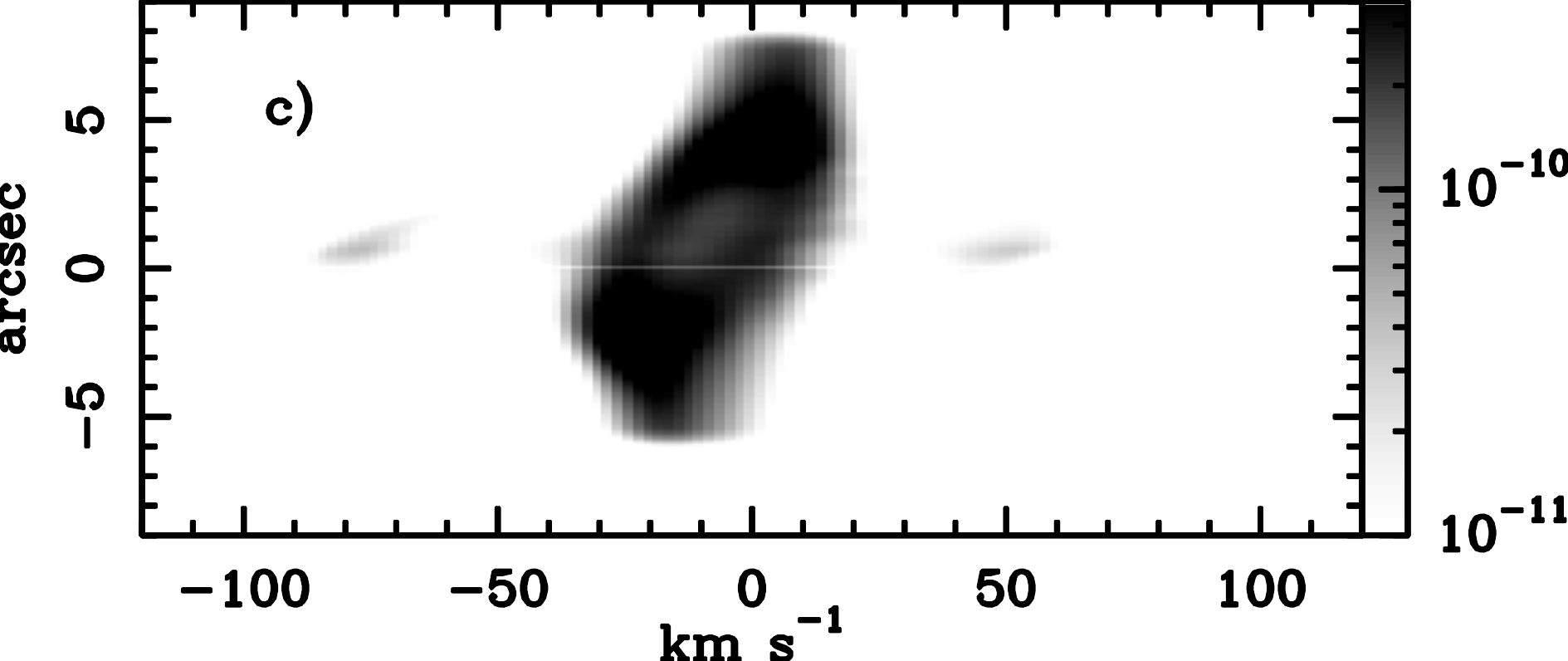}

\caption{Synthetic [N~II] PV diagrams for M\,3-15. a) PV diagram for slit located at 2.0
  arcsec to the right of the source. b) PV diagram for a slit
  going through the centre of the source. c) PV diagram for a slit at 2.0 arcsec to the
  left from the source. The horizontal and vertical axes represent the radial velocity
  in km s$^{-1}$ and the spatial direction in arcsec, respectively. The logarithmic grey scale is given in units of erg cm$^{-3}$ sr$^{-1}$.}
\label{fig:pv-M3-15}
\end{figure}

M\,3-15 was modeled with the same hydrodynamical setup as M\,1-32: a
dense torus and a conical jet, but using different parameters. By
comparing the images of M\,1-32 and M\,3-15 (see Figs. 2 and 3), we
note that M\,3-15 exhibits an elongation in the SE-NW
direction. This fact suggests that the jet axis (the
$z$-axis of the computational domain)
must be tilted with respect to the North-South direction
and to the line of
sight. The synthetic PV diagrams displayed in
Fig. \ref{fig:pv-M3-15}, were obtained after an integration
  time of 3500 yr, and after applying
two rotations to the computational domain. After several tests, we obtained
a good agreement with observations if the jet
axis was tilted by -55$^{\circ}$ with respect to the North direction,
 and at an
inclination of 30$^{\circ}$ with respect to the line of sight. These
PV diagrams show a bright condensation
corresponding to the torus. The
collimated outflows have velocities of the order of 100 km s$^{-1}$.

Our hydrodynamical
model, shown in Fig. \ref{fig:model}, has a bipolar structure of conical jets
which emerge through the poles of a toroidal structure.
Morphologically, M1-32 has been classified as ``elliptical"
\citep{Stanghellini2010},
and M3-15 as L,c \citep{SahaiMorris2011}, meaning that it
shows close, collimated lobes, with a ``barrel-shaped" central
region. In the observations, these objects do not look like bipolar nebulae due
to the bipolar lobes are being ejected along or very close to the line
of sight. Their bipolarity is only noticed through the spectroscopic
data. Therefore, we propose to call them ``spectroscopic
bipolars".

\begin{figure}
\centering
\includegraphics[width=1.05\linewidth]{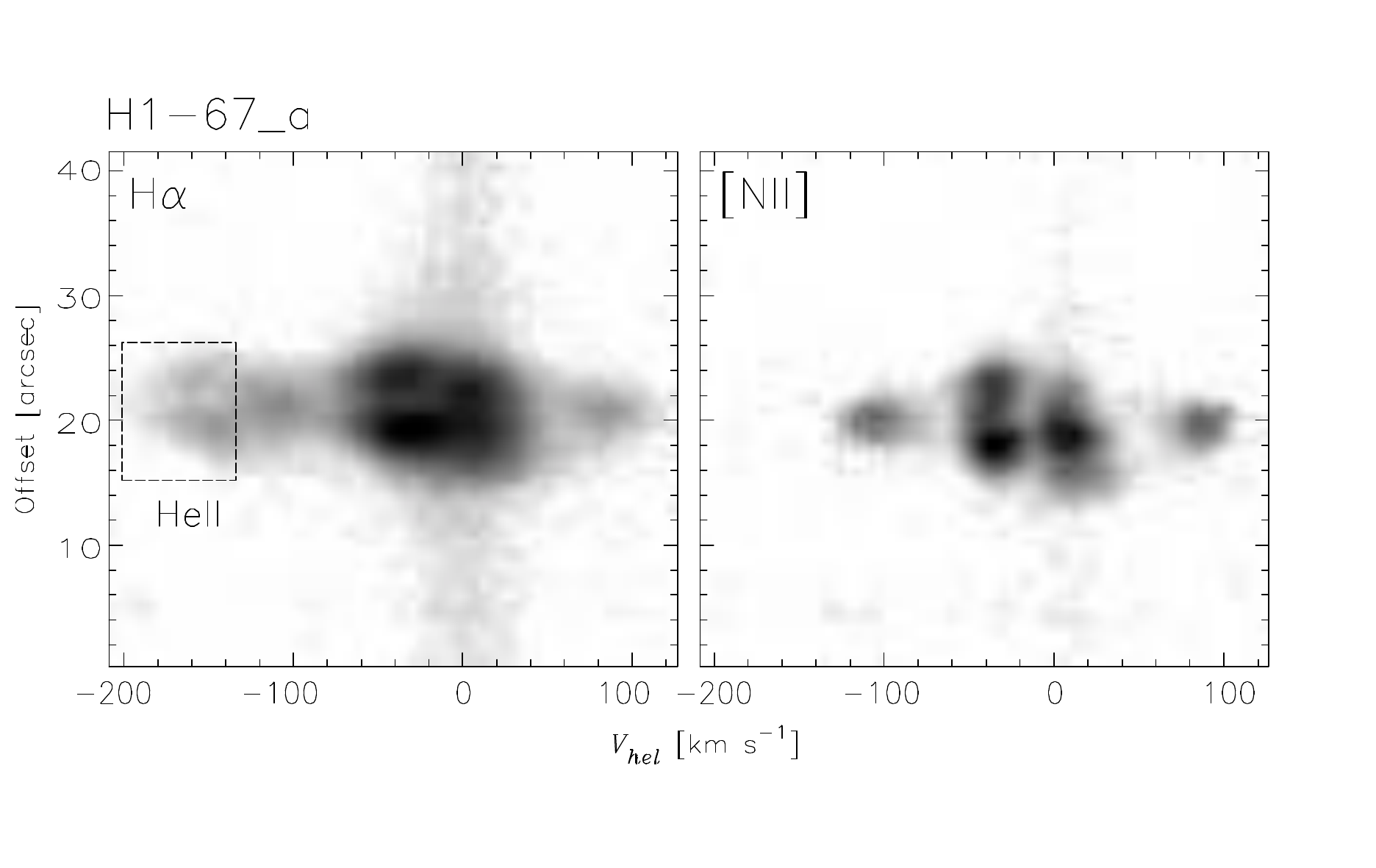}
\includegraphics[width=1.05\linewidth]{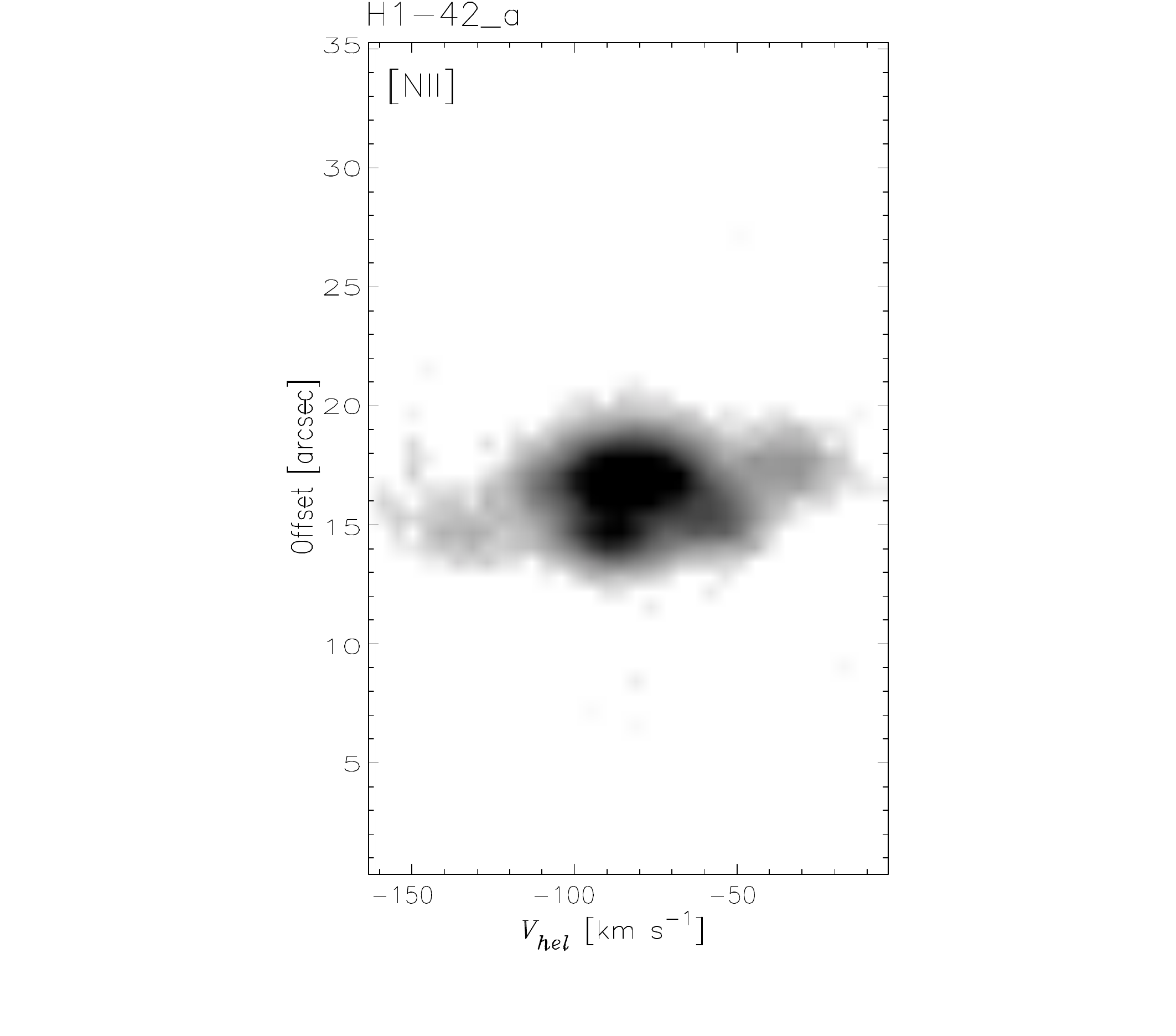}
\includegraphics[width=0.75\linewidth]{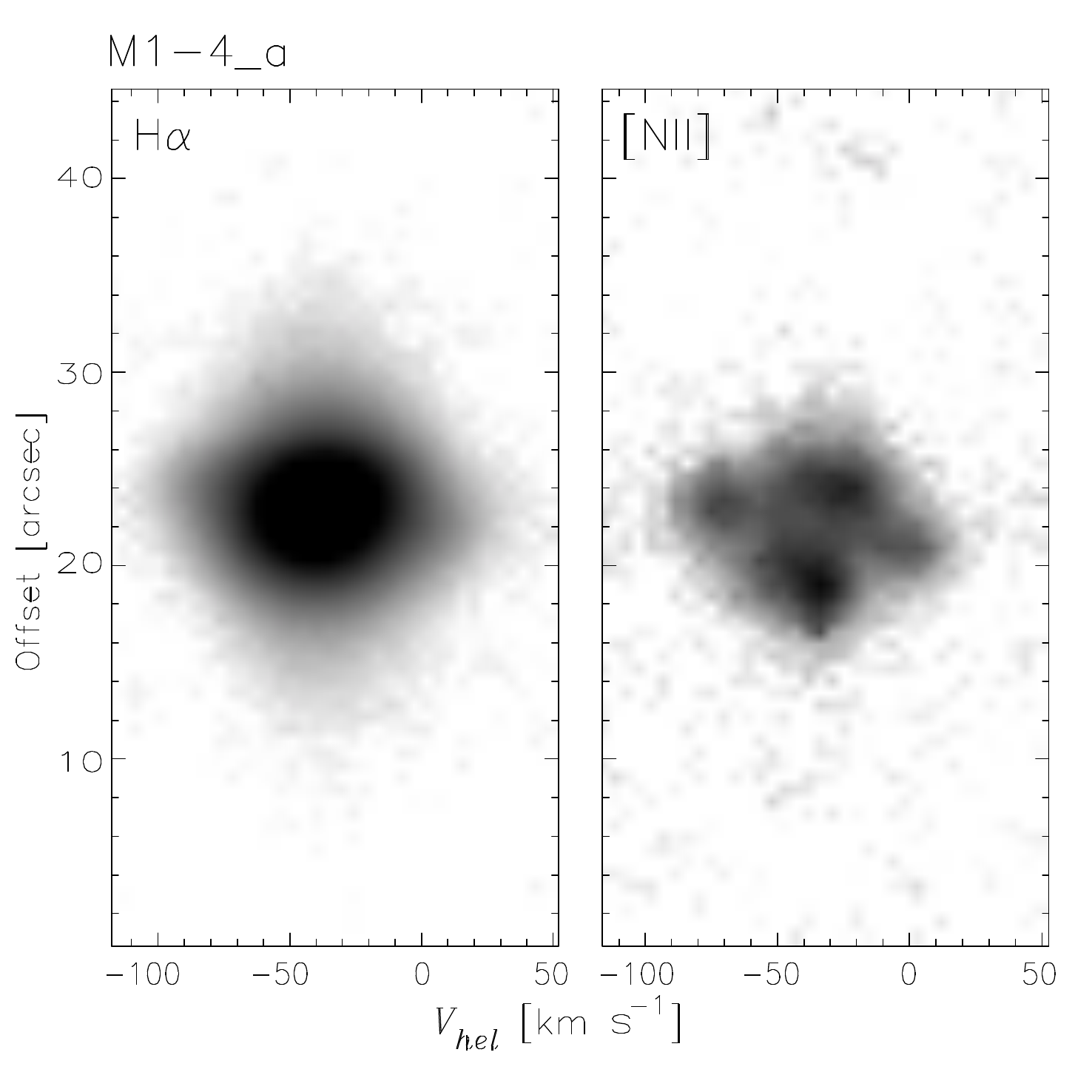}
\caption{PV diagrams observed for three PNe, obtained from The SPM
  Kinematic Catalogue of Planetary Nebulae.  Top: PN H\,1-67 with a [WO2] central star, we observe three bright knots at the centre and two faint knots reaching high velocities. Middle:  PN H\,1-42 with
  a WEL central star, we observe a sphere at the centre and high-velocity
  outflows.
  Bottom: PN M\,1-4 with a normal central star, we observe two
  bright, inclined knots at the centre, and two faint knots reaching low velocities. These PNe are good
  candidates for our hydrodynamical
  model.}
\label{fig:JETS}
\end{figure}

Using the available spectra in The SPM Kinematic Catalogue of Galactic
Planetary Nebulae \citep{Lopez2012}, we have found some PNe with
similar spectral characteristics to M\,1-32 and M\,3-15 (spectra
showing outflows).  From a sample of 29 objects with [WC] central
stars, 10 [WC]PNe are ``spectroscopic bipolar nebulae", that represents 34.4\%
of the sample. The outflows of these [WC]PNe reach more than 70 km
s$^{-1}$. As an example, in Fig. \ref{fig:JETS} we present the PV diagram for
H\,1-67 which has a [WO\,2] central star. Its outflows have a velocity of $\pm$ 100
km s$^{-1}$ relative to the system. 

From a sample of 34 PNe (taken
randomly) with WEL central stars (weak emission line stars), 7 objects
have outflows. This represents 20.5\% of the sample. Only three of
these seven objects (H\,1-42, NGC\,6644 and M\,3-38) would be ``spectroscopic
bipolar nebulae" with fast bipolar outflows at about 70 km
s$^{-1}$. A PV diagram for H\,1-42 is shown in Fig. \ref{fig:JETS}. These three PNe and 10 [WC]PNe mentioned above are good candidates  for modelling with our ``torus+bipolar outflow'' dynamical
model.

In addition, from a sample of 36 PNe with normal central stars (taken randomly), 4 have outflows.
These outflows show velocities smaller than 50 km s$^{-1}$. From these four, M\,1-4
 could be explained with the hydrodynamical model presented here but with a slower
velocity jet (see Fig. \ref{fig:JETS}).\\ 
According to these raw statistics, an important fraction of [WC] type
central stars  could be producing  PNe with high-velocity outflows, which is
not the case for PNe around  WELS or normal central stars.

\section{Conclusions}

With the Yguaz\'u code we have carried out 3D hydrodynamical simulations of a
bipolar jet system moving inside an anisotropic AGB wind with a high density
at the Equator (producing a dense torus around the
central star of a PN). In order to explain the global morphology and
the PV diagrams of the PNe M\,1-32 and M\,3-15, it was necessary to
assume that all of the simulated structure is photoionized by the central star.

The bipolar ejection leaves the central star, expanding
through the poles of the torus  at high velocities, of the
order of 160 km s$^{-1}$ and 100 km s$^{-1}$ for M\,1-32 and M\,3-15,
respectively.
With the simulations we reproduce the observed PV diagrams of these
nebulae after an integration time of 4500 yr for M\,1-32, and 3500 yr for
M\,3-15. These values are in agreement with the age of young PN.

\citet{SahaiMorris2011}
classified M\,3-15 as having close collimated lobes with a close barrel shaped central region,
and M\,1-32 was previously classified as an elliptical.
However, spectroscopically we find that both M\,3-15 and M\,1-32 show a torus and high-velocity bipolar
ejections. Therefore, we propose to call them ``spectroscopic bipolar nebulae".
As we explained in the introduction bipolar (and multipolar) planetary nebulae
can be created by binary stars \citep[see for
  example,][]{Morris1987, Soker1998}. Therefore, M\,1-32
and M\,3-15 could host a binary system in their centre, which could
be confirmed in a future work.

In the kinematical catalogue of Galactic PNe \citep{Lopez2012} we have
found that the spectra of 34.4\% of 29 [WC]PNe show outflows at
high velocities.  Some PNe around WELS or normal central stars also
show similar outflows, but with smaller velocities.  Therefore, a large fraction of 
[WC] central stars are producing nebulae with fast, bipolar outflows.

In summary, we have shown that the same physical model, a bipolar jet system moving
inside an anisotropic AGB wind (with high density at the Equator, producing
a dense torus-like structure) can explain
the observed PV diagrams of M\,1-32 and M\,3-15 PNe. In future work we intend
to model other PNe with similar characteristics.

\section*{Acknowledgments}
We would like to thank Fernando \'Avila Castro and Dr. Margarita
Pereyra for their help in the installation and use on WIP. Helpful comments by Dr. Jorge Garc\'ia-Rojas
are deeply acknowledged. We are grateful to Prof. Noam Soker for
his comments and suggestions, which allowed to improve the previous
version of this manuscript. We acknowledge Enrique Palacios (C\'omputo-ICN) for mantaining
the Linux servers where the hydrodynamical simulations were carried out. We thank the daytime 
and night support staff at the OAN-SPM for facilitating and helping to obtain our observations. This work
received finantial support from DGAPA-PAPIIT grants IN109614, IG100516
and CONACyT grant 167611. J.S.R.-G. acknowledges scholarship from
CONACyT-M\'exico.

\bibliographystyle{mn2e}

\label{lastpage}

\appendix

\end{document}